\documentclass[aps,prd,superscriptaddress,floatfix,nofootinbib,preprintnumbers,eqsecnum,twocolumn]{revtex4-2}

\pdfoutput=1
\usepackage{amssymb,amsfonts,amsmath,graphicx,bbm}
\usepackage{comment}
\usepackage{xcolor}
\usepackage{orcidlink}
\usepackage[english]{babel}
\usepackage[T1]{fontenc}
\usepackage{amsmath}
\usepackage{slashed}
\usepackage{booktabs}
\usepackage{listings}
\usepackage[utf8]{inputenc}
\usepackage{circuitikz}
\usepackage{nicematrix}
 \hypersetup{ 
   colorlinks,
   linkcolor={blue!80!black},
   citecolor={blue!70!black},
   urlcolor={blue!70!black}
 }
\usepackage{mathtools}
\usepackage{footmisc}
\usepackage{mathrsfs}
 
\usepackage{placeins}

\usepackage{amssymb,amsmath,epsfig,longtable}

\numberwithin{equation}{section}

\newcommand{\bean}{\begin{eqnarray*}}
\newcommand{\eean}{\end{eqnarray*}}

\newcommand{\fref}[1]{Figure~\ref{#1}}

\newcommand{\IP}{\mathbb{P}}

\newcommand{\cN}{{\cal N}}

\newcommand{\cA}{{\cal A}}
\newcommand{\cB}{{\cal B}}
\newcommand{\cC}{{\cal C}}
\newcommand{\cL}{{\cal L}}
\newcommand{\cK}{{\cal K}}
\newcommand{\cV}{{\cal V}}

\newcommand{\FS}{({\rm ref})}

\def\cjn1{{\cA, \cC^*\otimes \wedge^j \cN^*}}
\def\bjn1{{\cA, \cB^*\otimes \wedge^j \cN^*}}
\def\vjn1{{\cA, \cV^*\otimes \wedge^j \cN^*}}
\def\cjn2{{\cA, \cC\otimes \wedge^j \cN^*}}
\def\bjn2{{\cA, \cB\otimes \wedge^j \cN^*}}
\def\vjn2{{\cA, \cV\otimes \wedge^j \cN^*}}

\newcommand{\varstr}[2]{\vrule height #1 depth #2 width0pt}

\newcommand{\be}{\begin{equation}}
\newcommand{\ee}{\end{equation}}
\newcommand*{\nnbe}{\begin{equation}}
\newcommand*{\nnee}{\end{equation}}
\newcommand{\bea}{\begin{eqnarray}}
\newcommand{\eea}{\end{eqnarray}}
\newcommand{\ba}{\begin{align}}
\newcommand{\ea}{\end{align}}
\newcommand{\bi}{\begin{itemize}}
\newcommand{\ei}{\end{itemize}}

\newsavebox{\overlongequation}

\usepackage{hyperref}

\begin{document}
\title{Computation of Quark Masses from String Theory}


\author{Andrei Constantin}
\email[]{andrei.constantin@physics.ox.ac.uk}
\affiliation{Rudolf Peierls Centre for Theoretical Physics, University of Oxford, Parks Road, Oxford OX1 3PU, UK}

\author{Cristofero S. Fraser-Taliente}
\email[]{cristofero.fraser-taliente@physics.ox.ac.uk}
\affiliation{Rudolf Peierls Centre for Theoretical Physics, University of Oxford, Parks Road, Oxford OX1 3PU, UK}

\author{Thomas R. Harvey}
\email[]{thomas.harvey@physics.ox.ac.uk}
\affiliation{Rudolf Peierls Centre for Theoretical Physics, University of Oxford, Parks Road, Oxford OX1 3PU, UK}

\author{Andre Lukas}
\email[]{lukas@physics.ox.ac.uk}
\affiliation{Rudolf Peierls Centre for Theoretical Physics, University of Oxford, Parks Road, Oxford OX1 3PU, UK}

\author{Burt Ovrut}
\email[]{ovrut@physics.upenn.edu}
\affiliation{Department of Physics, University of Pennsylvania, Philadelphia, PA 19104, USA}


\begin{abstract}\noindent
We present a numerical computation, based on neural network 
techniques, of the physical Yukawa couplings in a heterotic string theory compactification on a smooth Calabi-Yau threefold with non-standard embedding.
The model belongs to a large class of heterotic line bundle models that have previously been identified and whose low-energy spectrum precisely matches that of the MSSM plus fields uncharged under the Standard Model group. 
The relevant quantities for the calculation, that is, the Ricci-flat Calabi-Yau metric, the Hermitian Yang-Mills bundle metrics and the harmonic bundle-valued forms, are all computed by training suitable neural networks. For illustration, we consider a one-parameter family in complex structure moduli space. The computation at each point along this locus takes about half a day on a single twelve-core CPU. Our results for the Yukawa couplings are estimated to be within 10\% of the expected analytic result. We find that the effect of the matter field normalisation can be significant and can contribute towards generating hierarchical couplings. We also demonstrate that a zeroth order, semi-analytic calculation, based on the Fubini-Study metric and its counterparts for the bundle metric and the bundle-valued forms, leads to roughly correct results, about 25\% away from the numerical ones. The method can be applied to other heterotic line bundle models and generalised to other constructions, including to F-theory models.
\begin{center}
{\bf Dedicated to the memory of Graham G.~Ross}
\end{center}
\end{abstract}

\maketitle		


\section{Introduction}
Computing the values of the  quark and lepton masses and understanding their hierarchical structure from first principles is a long-standing and fundamental open problem in theoretical particle physics. Arguably, string theory  currently provides the only framework which allows for such a computation.
However, the low-energy particle content obtained from string compactification was, for a long time, not sufficiently realistic to warrant detailed computations of couplings. Now that many models with the Standard Model spectrum are available, particularly in the context of $E_8\times E_8$ heterotic string compactifications on Calabi-Yau (CY) threefolds with holomorphic, poly-stable vector bundles \cite{Candelas:1985en,Braun:2005nv,Blumenhagen:2006ux,Blumenhagen:2006wj, Braun:2009qy,Braun:2011ni,Anderson:2011ns,Anderson:2012yf,Anderson:2013xka,Buchbinder:2013dna,Buchbinder:2014qda,Buchbinder:2014sya, Constantin:2018xkj, Constantin:2021for, Abel:2021rrj, Abel:2023zwg, Ambroso:2010pe,Ovrut:2015uea}, computing Yukawa couplings and the resulting fermion masses and mixing angles is the obvious next step.

The physical Yukawa couplings are completely specified by two moduli-dependent quantities in the low-energy four-dimensional $N = 1$ supersymmetric Lagrangian: the holomorphic Yukawa couplings, which arise in the superpotential, and the matter field Kähler metric which determines the normalisations of the $N=1$ chiral superfields. The holomorphic Yukawa couplings, defined in Eq.~\eqref{eqn:HolYuk}, are quasi-topological and do not depend on the Ricci-flat metric of the CY threefold. Consequently, they can often be calculated analytically using algebraic or differential geometric tools. This has been carried out for a number of  models~\cite{Braun:2006me, Anderson:2010tc, Blesneag:2015pvz,Blesneag:2016yag,Blesneag:2021wdf}. On the other hand, the computation of the matter field Kähler metric, defined in Eq.~\eqref{eqn:MFM},    requires full knowledge of the compactification geometry, including the Ricci-flat CY metric, the Hermitian Yang-Mills (HYM) connection on the holomorphic vector bundle, as well as various harmonic bundle-valued forms on the CY threefold.
Obtaining these quantities has been the single major hurdle (other than finding models with a realistic particle spectrum) in the computation of Yukawa couplings from string theory for nearly four decades, as no closed-form expressions are known.

A salient exception to this rule is represented by standard embedding CY compactifications of the $E_8\times E_8$ heterotic string (and singular limits thereof, such as toroidal orbifold compactifications) \cite{Strominger:1985ks, Candelas:1987se, Dixon:1989fj, Candelas:1990pi, Ishiguro:2021drk}. In this setting, the matter field K\"ahler metric and the holomorphic Yukawa couplings can be explicitly calculated from period integrals via special geometry or by using conformal field theory techniques. In fact, the pioneering work on fermion masses from string theory~\cite{greene:1986ar,greene:1986bm,greene:1986jb,greene:1987xh} was carried out in this context. These analytic results for models with standard embedding have recently been successfully reproduced, and extended, using neural networks methods~\cite{Butbaia:2024tje}. Unfortunately, the class of standard embedding models is very limited in terms of what can be achieved, both at the level of the spectrum and couplings. 

The methods presented in this paper apply to a wide class of phenomenologically attractive models relying on line bundle sums over smooth CY threefolds embedded in products of projective spaces with freely acting discrete symmetries~\cite{Anderson:2011ns,Anderson:2012yf,Anderson:2013xka}, and may be further extended to compactifications with non-Abelian bundles. The core idea is to use neural networks to approximate the above-mentioned geometrical quantities, which are obtained as solutions of certain partial differential equations. As zeroth order 'reference' quantities (and for benchmarking) we use analytic expressions for the Fubini-Study K\"ahler form, the Chern connection and various bundle-valued differential forms, defined on the ambient space and subsequently restricted to the CY manifold, which represent the correct cohomology classes. To these reference quantities we add terms exact in cohomology which are represented by neural networks and optimised in order to obtain the Ricci-flat CY metric, the HYM bundle metrics and the harmonic forms\footnote{This is not to say we are computing the leading order correction in some expansion. The final results are precise, up to numerical error.}. The optimisation process involves sampling points on the CY threefold and minimising a loss function that takes into account how well the associated partial differential equations and various patching conditions are satisfied. Our methods build directly on the tools developed in the {\sf cymetric} package~\cite{Larfors:2021pbb,Larfors:2022nep} for the construction of numerical CY metrics, as well as earlier numerical work on:  
%
(a) numerical methods for of Ricci-flat CY metrics based on Donaldson's algorithm~\cite{donaldson2005numerical, Braun:2007sn, Douglas:2006rr}, functional minimisation~\cite{Headrick:2009jz}, and machine learning techniques~\cite{Ashmore:2019wzb,Anderson:2020hux, Jejjala:2020wcc, Douglas:2020hpv,Larfors:2021pbb,Larfors:2022nep,Ashmore:2021ohf,Gerdes:2022nzr}; (b) numerical methods for the computation of Hermitian Yang-Mills connections \cite{MR2154820,Douglas:2006hz,Anderson:2011ed, Anderson:2010ke,Ashmore:2021rlc,Ashmore:2023ajy}; (c) numerical harmonic functions and harmonic bundle-valued forms \cite{Braun:2008jp, Ashmore:2020ujw, Ashmore:2021qdf, Ashmore:2023ajy}. 

The main goal of this paper is to develop the neural-network approach up to the point where explicit values for the quark and lepton masses can be computed for arbitrary values of the moduli fields. In practice,
we carry out the computation of perturbative up-quark Yukawa couplings in a specific heterotic line bundle model, originally introduced in Refs.~\cite{Buchbinder:2014qda,Buchbinder:2013dna}, thus proving that such calculations are now feasible. 
A systematic exploration of the moduli space for this and other models is beyond the scope of the present letter, but likely within reach. The aim of this future work is to identify concrete line bundle models and specific loci in their moduli space that give rise to the observed flavour structure of the Standard Model. Combining such an analysis with moduli stabilisation may lead to a comprehensive explanation of the parameters in the Standard Model. 

It is worth pointing out that the present work does not rely on any simplifying assumptions, such as localisation, the only limiting factor being numerical accuracy. Localisation techniques have been heavily used in other contexts, including calculations of zero-mode wavefunctions on toroidal backgrounds \cite{Cremades:2004wa, Krippendorf:2010hj}, for F-theory models~\cite{Font:2008id, Heckman:2008qa, Hayashi:2009ge, Cecotti:2009zf, Font:2009gq, Conlon:2009qq, Hayashi:2009bt, Aparicio:2011jx, Palti:2012aa} and for heterotic models~\cite{Blesneag:2018ygh}. For compactifications on non-flat spaces, localisation relies on the observation that sufficiently large fluxes lead to localised matter field wave functions, so that approximate calculations can be carried out with a (locally) flat metric. This appears to circumvent the need to know the Ricci-flat CY metric explicitly. However, this method comes with a number of problems: it is difficult to assess its accuracy and to express the results in terms of standard CY moduli and there is a tension between the large fluxes required for localisation and the requirements of three families and anomaly cancellation. 

The structure of the paper is as follows. We begin in Section~\ref{sec:model} by reviewing the details of the heterotic string model under consideration, before moving on to describing the mathematical background and the computational setup in Section \ref{sec:training}. In Section~\ref{sec:MassesMixings}, we present our results for the physical up-quark Yukawa couplings and resulting fermion masses. We conclude in Section \ref{sec:Conc}. Further details of the calculation will be presented in a forthcoming longer paper \cite{futurePaper}.

\section{The heterotic string model}\label{sec:model}
The model underlying our computations was originally introduced in Ref.~\cite{Buchbinder:2014qda,Buchbinder:2013dna}. It is based on compactifying the $E_8\times E_8$ heterotic string on a smooth $\Gamma=\mathbb{Z}_2\times\mathbb{Z}_2$ quotient of a CY hypersurfaces $X$ of multi-degree $(2,2,2,2)$ in a product of four $\IP^1$-spaces. These hypersurfaces are, from now on, referred to as tetra-quadric CY threefolds or TQ threefolds, for short. The quotient $X/\Gamma$ has four K\"ahler parameters and $20$ complex structure parameters. The $\mathbb{Z}_2\times\mathbb{Z}_2$ symmetry descends from the ambient space~\cite{Braun:2010vc}, where it is generated by the matrices
\begin{equation}\label{Z2Z2def}
 \left(\begin{array}{rr}1&\!\!\!0\\0&\!-1\end{array}\right)\; ,\quad \left(\begin{array}{rr}0&~1\\1&~0\end{array}\right)\; ,
\end{equation}
acting simultaneously on the homogeneous coordinates of each $\IP^1$.

The standard K\"ahler forms on the four $\mathbb{P}^1$ factors, restricted to $X$, provide a basis, $(J_i)$, where $i=1,2,3,4$, for the second cohomology of $X$. The first Chern class of a line bundle $\cL\rightarrow X$ can be written as $c_1(\cL)=k^iJ_i$, where $k^i\in\mathbb{Z}$, and such a line bundle is also denoted by $\cL={\cal O}_X({\bf k})$, with ${\bf k}=(k^1,k^2,k^3,k^4)^T\in\mathbb{Z}^4$.

The vector bundle $V\rightarrow X$ is chosen to be a sum of five line bundles, that is $V=\bigoplus_{a=1}^5\cL_a$, with Chern classes $c_1(\cL_a)=k_a^iJ_i$. This bundle breaks one of the two $E_8$ gauge factors to $SU(5)\times S(U(1)^5)$. For the model under considerations, the five line bundles are specified by the column vectors of the matrix
\begin{equation}\label{Li}
\begin{array}{rcl}
  &&\quad\begin{array}{rrrrr}\cL_1\,& \cL_2\,& \cL_3\,& \cL_4\,& \cL_5\end{array}\\[2mm]
  ({k^i}_a)&=&\left[\begin{array}{rrrrr}
  -1&-1&0&1&1\\
  0&-3&1&1&1\\
  0&2&-1&-1&0\\
  1&2&0&-1&-2
  \end{array}\right]\;.
\end{array}  
\end{equation} 
The additional $U(1)$-symmetries are Green-Schwarz anomalous and their associated gauge bosons are super-heavy. At low energy they appear as global symmetries which constrain the allowed couplings. Finally, the $SU(5)$-gauge symmetry is broken to the Standard Model gauge group by specifying a discrete Wilson line with structure group $\Gamma=\mathbb{Z}_2\times\mathbb{Z}_2$. The model is free of gauge and gravitational anomalies (by a suitable choice of a five-brane or hidden bundle) and whatever remains from the second $E_8$ gauge group at low energy is `hidden', in the sense that all observable fields are uncharged under it. It is also supersymmetric along the locus where all K\"ahler moduli take the same value, that is,
\begin{equation}\label{t_locus}
t:=t_1=t_2=t_3=t_4\; .
\end{equation}
The particle content is that of the MSSM, plus a number of singlet fields uncharged under the SM gauge group. These fields are decorated with $U(1)$ charges and given~by
\begin{equation}\label{spectrum}
\begin{aligned}
   \begin{array}{c c c}
   2\,Q_2, 2\,U_2, 2\,E_2 \leftrightarrow \cL_2 & ~~~~&  Q_5, U_5, E_5 \leftrightarrow \cL_5\\[2pt]
  2\,D_{2,5}, 2\,L_{2,5} \leftrightarrow \cL_4\otimes \cL_5 & ~~~~& D_{2,4}, L_{2,4}\leftrightarrow \cL_2\otimes \cL_4\\[2pt]
   H^d_{2,5} \leftrightarrow \cL_2\otimes \cL_5&~~~~&
   H^u_{2,5} \leftrightarrow \cL_2^\ast\otimes \cL_5^\ast\\[2pt]
   3\,S_{2,4} \leftrightarrow \cL_2\otimes \cL_4^\ast &~~~~& \text{12 other singlets}
   \end{array}
\end{aligned}
\end{equation}
The subscripts label the $U(1)$ symmetries under which the particles carry charge $1$, while being uncharged under all other $U(1)$ symmetries. The exceptions are $H^u_{2,5}$, whose only non-zero charges are $-1$ under the second and fifth $U(1)$ symmetry, and $S_{2,4}$ with charge $+1$ under the second symmetry and charge $-1$ under the fourth\footnote{Note that the up and down Higgs triplets have been projected out by the $\mathbb{Z}_2\times\mathbb{Z}_2$ quotient and the inclusion of the Wilson line.}. The fields are in one-to-one correspondence with certain harmonic bundle-valued one-form on specific line bundles, which have been indicated in Eq.~\eqref{spectrum}.

The $U(1)$ symmetries enforce the vanishing of  down-quark and lepton Yukawa matrices at the perturbative level; for a realistic model, they would have to be generated non-perturbatively. 
Writing the left-handed quarks as $(Q^i)=(Q_2^1,Q_2^2,Q_5)$, the right-handed up-quarks as $(U^i)=(U_2^1,U_2^2,U_5)$ and the up-Higgs as $H^u=H^u_{2,5}$, the  holomorphic up-quark Yukawa couplings are of the form
\begin{equation}
 W_{\rm u} = Y^u_{ij}H^uQ^iU^j~,
\end{equation} 
where $i,j=1,2,3$ label the three quark families. The $U(1)$ symmetries enforce a specific   
structure of the up-Yukawa matrix given by
\begin{equation}\label{Yhol}
    Y^u = \left(\begin{array}{ccc}
        0 & 0& \lambda_1 \\
        0 & 0& \lambda_2 \\
        \lambda_3 & \lambda_4& 0 
    \end{array} \right)~.
\end{equation}
Its entries $\lambda_1,..,\lambda_4$ are quasi-topological and can be computed using differential geometric techniques, as detailed in Refs.~\cite{Blesneag:2015pvz,Blesneag:2016yag,Blesneag:2021wdf}~\footnote{Unfortunately, there appears to be a mistake in the calculation carried out in Ref.~\cite{Blesneag:2015pvz} of the holomorphic Yukawa couplings for this model, due to a missed boundary term, an issue which we correct in the present paper.}. For completeness, we note that the full perturbative superpotential is 
\begin{equation}
W = W_{\rm u} +\rho_{\alpha i} S_{2,4}^{\alpha}L_{4,5}^{i} H_{2,5}^u, \label{Wsm}
\end{equation}
where the index $\alpha=1,2,3$ labels the three singlets $S_{2,4}$ present in the spectrum. These are interpreted as right-handed neutrinos. 

The part of the K\"ahler potential relevant for the calculation of the physical up-Yukawa couplings has the form
\[
K=K^Q_{ij}Q^i\bar{Q}^j+K^u_{ij}U^i\bar{U}^j+k H^u\bar{H}^u,
\]
and $U(1)$-invariance dictates the following structure for the K\"ahler metrics:
\begin{equation}\label{Kstruct_sec2}
    K^Q=\mathcal{V}^{-\frac{1}{3}}\begin{pNiceArray}{ccc}
        \Block{2-2} {~~\cK^Q}  & &0\\
        & & 0\\
        0&0&k^Q
        \end{pNiceArray},\;
    K^u=\mathcal{V}^{-\frac{1}{3}}\begin{pNiceArray}{ccc}
        \Block{2-2} {~~\cK^u}  & &0\\
        & & 0\\
        0&0&k^u
        \end{pNiceArray}.\;
\end{equation}  
The factor $\cV^{-1/3}$ captures the full K\"ahler moduli dependence in this model due to Eq.~\eqref{t_locus}. This means the complex $2\times 2$ matrices $\cK^Q$, $\cK^u$ and the real numbers $k$, $k^Q$, $k^u$ in Eq.~\eqref{Kstruct_sec2} are K\"ahler moduli independent, but they still depend on complex structure.
After bringing the resulting kinetic terms into canonical form, one finds the physical up-Yukawa matrix
\begin{equation}\label{Lphys}
\begin{aligned} 
Y_{\rm phys}^u = & \left(\begin{array}{ccc}
        0 & 0& a_1 \\
        0 & 0& a_2 \\
        b_1&b_2&0
    \end{array} \right)~\\
\begin{pNiceArray}{c}a_1\\a_2\end{pNiceArray}=\frac{e^{-\phi}}{\sqrt{kk^u}}P_Q \begin{pNiceArray}{c}\lambda_1\\ \lambda_2\end{pNiceArray},&\quad 
 \begin{pNiceArray}{c}b_1\\b_2\end{pNiceArray} =\frac{e^{-\phi}}{\sqrt{kk^Q}}P_u \begin{pNiceArray}{c}\lambda_3\\\lambda_4\end{pNiceArray},
\end{aligned}
\end{equation}
where $\phi$ is the dilaton, and $P_Q$ and $P_u$ are $2\times 2$ diagonalising matrices satisfying
\begin{equation}\label{KP}
P_Q\cK^QP_Q^\dagger=\mathbbm{1}_2\;,\quad P_u\cK^uP_u^\dagger=\mathbbm{1}_2\;.
\end{equation}
Note that the volume dependence of the physical Yukawa couplings drops out due to the additional factor of $e^{-\phi}/\sqrt{\mathcal V}$ which arises from the $\exp(K/2)$ prefactor to the Yukawa couplings in the component supergravity Lagrangian.\footnote{Note that this is a general feature. The physical Yukawa couplings in heterotic theories are independent of the overall CY volume modulus.}
Finally, the up-quark masses are 
\begin{equation}\label{masses}
    (m_1,m_2,m_3)=
    |\langle H^u\rangle| e^{-\phi} \Bigg(0,\frac{\left|P_Q\begin{pNiceArray}{c}\lambda_1\\ \lambda_2\end{pNiceArray}\right|}{\sqrt{kk^u}},\frac{\left|P_u \begin{pNiceArray}{c}\lambda_3\\ \lambda_4\end{pNiceArray}\right|}{\sqrt{kk^Q}}\Bigg)\; .
\end{equation}
A number of comments are in order at this point. 
Firstly, in this model, the holomorphic up-quark Yukawa matrix~\eqref{Yhol} on its own cannot lead to a non-zero mass for the first generation due to its reduced rank. A non-zero up-quark mass would have to be generated non-perturbatively. Secondly, a potential split between the second and third generation up-quark masses can be induced by the structure of the holomorphic up-Yukawa matrix as well as by the non-canonical matter field Kähler metric, both of which depend on complex structure moduli. The dependence of the physical Yukawa couplings on the K\"ahler moduli completely drops out in this particular model, as a consequence of the volume-independence mentioned earlier, and of working at the special locus Eq.~\eqref{t_locus} in K\"ahler moduli space. Thirdly, in order to make contact with the measured values of the quark masses, one should include the RG running from the compactification scale down to the electroweak scale in the presence of supersymmetry breaking. This is amenable to standard methods and will not be discussed further. 
Finally, loop, $\alpha'$, and non-perturbative corrections can affect the physical Yukawa couplings. These are suppressed in the large volume and weak coupling regime and, in these limits, will only lead to small corrections to non-zero perturbative masses (of course, they may be the leading effect if a perturbative mass vanishes, as in our present example). For precise predictions from string theory, all these corrections must ultimately be considered. 

\section{Metrics and harmonic forms}\label{sec:training}
In geometric heterotic compactifications on smooth CY threefolds $X/\Gamma$, the holomorphic Yukawa couplings and the matter field K\"ahler metric are given by the following expressions:
\begin{align}
    \label{eqn:HolYuk}
    \lambda_{IJK} &= \gamma_{IJK} \frac{c \, 2\sqrt{2}}{|\Gamma|} \int_X \nu_I \wedge \nu_J \wedge \nu_K \wedge\Omega,\\
    \label{eqn:MFM}
    K_{IJ} &=\frac{\zeta_{IJ}}{2\mathcal{V} |\Gamma|}\int_X \nu_I \wedge\star_V \nu_J = \frac{\zeta_{IJ}}{2\mathcal{V} |\Gamma|}\int_X \nu_I \wedge\star (H_{J} \bar{\nu}_J)\; .
\end{align}
Here $\nu_I$ are harmonic $(0,1)$-forms which represent the matter fields and take values in line bundles $\cL_I$, whilst $H_I$ are HYM bundle metrics on $\cL_I$. Concretely, for the model outlined in the previous section, the line bundles $\cL_I$ are the ones given in Eq.~\eqref{spectrum}. Furthermore, $\mathcal{V}$ is the CY volume and the Hodge star is taken with respect to the Ricci-flat CY metric. The holomorphic $(3,0)$-form $\Omega$ on $X$ is normalised such that $\int_X\Omega\wedge\bar{\Omega}=1$. The integrals are performed on the `upstairs' manifold $X$ and the result is transferred to the smooth `downstairs' quotient $X/\Gamma$ by  dividing by the group order, $|\Gamma|$. Concretely, for our specific model, we have $|\Gamma|=|\mathbb{Z}_2\times\mathbb{Z}_2|=4$. In line with the constraints from the low-energy $U(1)$ symmetries, holomorphic Yukawa couplings $\lambda_{IJK}$ can be non-zero only if $\cL_I\otimes\cL_J\otimes \cL_K=\mathcal{O}_X$ and entries $K_{IJ}$ for $\cL_I\neq \cL_J$ must vanish. 
The numerical pre-factors arise from the dimensional reduction, whilst the factor $c=\sqrt{H_IH_JH_K}$ originates from transforming between conventions used in the physics and mathematics literature~\cite{Douglas:2006hz}. 
Note that $H_IH_JH_K$ is constant for any Yukawa coupling allowed by the $U(1)$ symmetries. The factors $\zeta_{IJ}$ and $\gamma_{IJK}$ are group theoretic factors, coming from the branching of the 10D $E_8$ gauge group. Their derivation will be given in the upcoming paper, and for the case of up-quark Yukawa couplings they are given by
\begin{equation}
    \zeta_{IJ} = \frac{\delta_{IJ}}{2} ,\,\, \gamma_{IJK} = \frac{1}{8\sqrt{30}}.
\end{equation}
We write $\gamma_{IJK}$ as a constant as it takes the same value, up to a sign, for all terms allowed by the symmetries. The signs are such to cancel the anti-symmetry of the forms inside~\eqref{eqn:HolYuk}.

As mentioned in the previous section, to compute the Ricci-flat CY metric $g$, the HYM bundle metrics $H_I$ and the harmonic forms $\nu_I$, we start with certain reference quantities which represent the correct cohomology classes. To these, we add exact terms which are determined by training suitable neural networks. We now describe how this is done for each of the three types of quantities in turn.

\subsection{The Ricci-flat CY metric}\label{sec:phi}
\subsubsection{Mathematical background}
Yau's theorem, applied to a CY manifold $X$, asserts that in any given K\"ahler class, associated to a reference metric $g^{({\rm ref})}$, there exists a unique Ricci-flat metric
\begin{equation}\label{phidef}
 g_{a\bar b} = g^{(\rm ref)}_{a\bar b} + \partial_a \bar\partial_{\bar b}\phi,
\end{equation}
where $\phi$ is a real function on $X$ determined by solving the relevant Monge-Amp\`ere equation. In practice, this can be done by training a neural network which represents $\phi$. This approach has been realised in the {\sf cymetric} package~\cite{Larfors:2021pbb,Larfors:2022nep}, where it is referred to as the `$\phi$-model'.

Our specific simply-connected CY threefold is defined as a tetra-quadric (TQ) hypersurface given as the zero locus of a defining polynomial $p$ of multi-degree $(2,2,2,2)$, in the ambient space $\mathcal{A}=\mathbb{P}^1\times\mathbb{P}^1\times\mathbb{P}^1\times\mathbb{P}^1$. Homogeneous coordinates on the four $\mathbb{P}^1$s are denoted by $x_\alpha,y_\alpha,u_\alpha,v_\alpha$, where $\alpha=0,1$. The standard patches on $\mathcal{A}$ with $x_\alpha,y_\beta,u_\gamma,v_\delta\neq 0$ are denoted by $U_{\alpha\beta\gamma\delta}$ and affine coordinates on the patch $U_{0000}$ are defined by $z_1=x_1/x_0$, $z_2=y_1/y_0$, $z_3=u_1/u_0$ and $z_4=v_1/v_0$. We also introduce the convenient shorthand $\kappa_i=1+|z_i|^2$. 

Computing the holomorphic Yukawa couplings~\eqref{eqn:HolYuk} requires the holomorphic $(3,0)$-form $\Omega$. On the standard patch with coordinates $z_i$ defined above, it can be written explicitly as
\begin{equation}\label{omega}
 \Omega \propto \hat\Omega=\left.\frac{dz_1\wedge dz_2\wedge dz_3}{\frac{\partial p}{\partial z_4}}\right|_X\;,
\end{equation}
with the proportionality constant fixed by $\int_X\!\Omega{\wedge}\bar{\Omega}=1$, up to an arbitrary phase that drops out of physical quantities.

For the reference metric in Eq.~\eqref{phidef} we choose the Fubini-Study metric restricted to $X$:
\begin{equation}
    g_{a\bar{b}}^{({\rm ref})}=\left.\sum_{i=1}^4\frac{t^i}{2\pi}\partial_a\bar{\partial}_{\bar{b}}\ln(\kappa_i)\right|_X\; ,
\end{equation}
where $t^i\in\mathbb{R}^{>0}$ are the four K\"ahler parameters. In terms of these parameters, the CY volume $\mathcal{V}$ reads:
\begin{equation}
\mathcal{V}=2(t_1t_2t_3+t_1t_2t_4+t_1t_3t_4+t_2t_3t_4)\; .
\end{equation}
At the supersymmetric locus~\eqref{t_locus}, this expression simplifies to $\mathcal{V}=8t^3$, with the overall K\"ahler parameter $t$.

For most of the calculations below, we will be working with the two-parameter family of TQs defined by the vanishing of the polynomial
\begin{equation}\label{pdef}
    p=\!\!\!\!\!\!\!\sum_{\substack{{\rm even}\\[2pt] \alpha{+}\beta{+}\delta{+}\gamma}}\!\!\!\!\!\!\!x_\alpha^2y_\beta^2u_\gamma^2v_\delta^2 +\psi_0\!\!\!\!\!\!\!\sum_{\substack{{\rm odd}\\[2pt] \alpha{+}\beta{+}\delta{+}\gamma}}\!\!\!\!\!\!\!x_\alpha^2y_\beta^2u_\gamma^2v_\delta^2+\psi\!\!\!\!\prod_{\alpha,\beta,\delta,\gamma}\!\!\!\! x_\alpha y_\beta u_\gamma v_\delta,
\end{equation}
constructed in analogy with the Dwork pencil of quintic threefolds. For $\psi_0\neq 1$, this leads to a smooth hypersurface for generic values of $\psi$. This polynomial is of course invariant under the action \eqref{Z2Z2def} of $\Gamma=\mathbb{Z}_2\times\mathbb{Z}_2$. However, it is also invariant under an additional symmetry which enforces equality between the two non-zero up-quark masses, a degeneration reflected in the numerical calculation below (see, for instance, \fref{fig:mainresult}). To~illustrate that this degeneracy can be lifted, we also consider a more general $\Gamma=\mathbb{Z}_2\times\mathbb{Z}_2$ invariant polynomial which breaks the additional symmetry present in Eq.~\eqref{pdef}, namely,
\begin{equation}\label{eqn:newPoly}\begin{aligned}
    \tilde p =&\, 
    6 x_{1}^2 y_{0} y_{1} u_{0} u_{1} v_{0}^2
     +x_{0}^2 y_{0} y_{1} u_{0} u_{1} v_{0}^2
    -18 x_{0}^2 u_{0} u_{1} v_{0} v_{1}
   y_{0}^2
   \\& -2 x_{0} x_{1} y_{1}^2 v_{0} v_{1}
   u_{0}^2
   +7 x_{1}^2 y_{0} y_{1} v_{0} v_{1}
   u_{0}^2
   -11 x_{0}^2 y_{0}^2 u_{0}^2
   v_{0}^2
   \\&-2 x_{0} x_{1} y_{0}^2 v_{0} v_{1}
   u_{0}^2
   +7 x_{1}^2 y_{1}^2 u_{1}^2
   v_{0}^2
   -12 x_{0}^2 y_{1}^2 u_{1}^2 v_{0}^2
   \\&-6 x_{0} x_{1} y_{0} y_{1} u_{1}^2 v_{0}^2
   +20 x_{1}^2 y_{0}^2 u_{1}^2 v_{0}^2
   +4 x_{0}^2 y_{0}^2 u_{1}^2 v_{0}^2
   \\&-13
   x_{0} x_{1} y_{1}^2 u_{0} u_{1} v_{0}^2
   -9
   x_{0} x_{1} y_{0}^2 u_{0} u_{1} v_{0}^2
   +x_{1}^2 y_{0}^2 u_{0}^2 v_{0}^2
   \\&-15
   x_{1}^2 y_{1}^2 u_{0}^2 v_{0}^2
   -4 x_{0}^2 y_{1}^2
   u_{0}^2 v_{0}^2
   -4 x_{0} x_{1} y_{0} y_{1} u_{0}^2
   v_{0}^2
   \\&
   -13 x_{0}^2 y_{0} y_{1} v_{0} v_{1}
   u_{0}^2
   + (1\leftrightarrow 0)\; .
\end{aligned}\end{equation}
Here, $(1\leftrightarrow 0)$ is the instruction to copy over all the previous terms with coordinate indices swapped as indicated. 

Throughout the entire calculation below, we will set the overall K\"ahler parameter to $t=1$. Recall that in the present model the physical Yukawa couplings are independent of the K\"ahler parameters, so this choice does not limit the scope of our calculation.

\subsubsection{Computational realisation}
An approximate Ricci-flat CY metric is constructed via Eq.~\eqref{phidef}, where the function $\phi$ is represented by a neural network and trained on a loss function that includes the Monge-Amp\`ere loss, the transition loss and the K\"ahler class loss: 
\begin{equation}\label{phiLoss}
    \begin{aligned}
        \mathscr{L}  &= \alpha_1 \mathscr{L}_\text{MA}+\alpha_2\mathscr{L}_\text{tr.}+\alpha_3\mathscr{L}_{\text{K\"ahler}}\\
\mathscr{L}_\text{MA}[\phi]&=\ {\Bigg |\Bigg | } 1- \frac{1}{\kappa}\frac{J(\phi)\wedge J(\phi)\wedge J(\phi)}{\hat\Omega\wedge\overline{\hat \Omega}} {\Bigg |\Bigg | }_1\\
\mathscr{L}_\text{tr.}[\phi]&=\sum_{s\neq t} \big|\big|\phi_s-\phi_t\big|\big|_1\; .
    \end{aligned}
\end{equation}
Here $\alpha_1$, $\alpha_2$, $\alpha_3$ are weights for the three contributions and $||\cdot||_1$ is the $L_1$-norm computed by performing Monte-Carlo integration on $X$ as detailed in Ref.~\cite{Larfors:2022nep}. Furthermore, $J(\phi)$ is the K\"ahler form associated to the metric~\eqref{phidef}, $\hat\Omega$ is defined in Eq.~\eqref{omega}, $\phi_s$ is the version of $\phi$ computed on the patch $U_s$, while the normalisation factor $\kappa$ and the K\"ahler class loss $\mathscr{L}_{\text{K\"ahler}}$ are defined in Ref.~\cite{Larfors:2022nep}.

We carry out the computation of the CY metric at the points along the pencil of tetra-quadrics \eqref{pdef} defined by 
$\psi_0=2$ and for $\psi\in\{0,0.5,1,2,4,6\}$. 
For each value of $\psi$, the {\sf cymetric} package~\cite{Larfors:2021pbb,Larfors:2022nep} is used to create a sample of $300,000$ points on $X$, distributed according to the measure defined in Ref.~\cite{Larfors:2022nep}. These points are used to train the neural network for $\phi$ as well as the neural networks discussed below and to perform Monte-Carlo integration over~$X$. We split the point sample into training and validation sets at a ratio of 9:1. 

\begin{figure}
    \centering
    \includegraphics[width=0.95\linewidth]{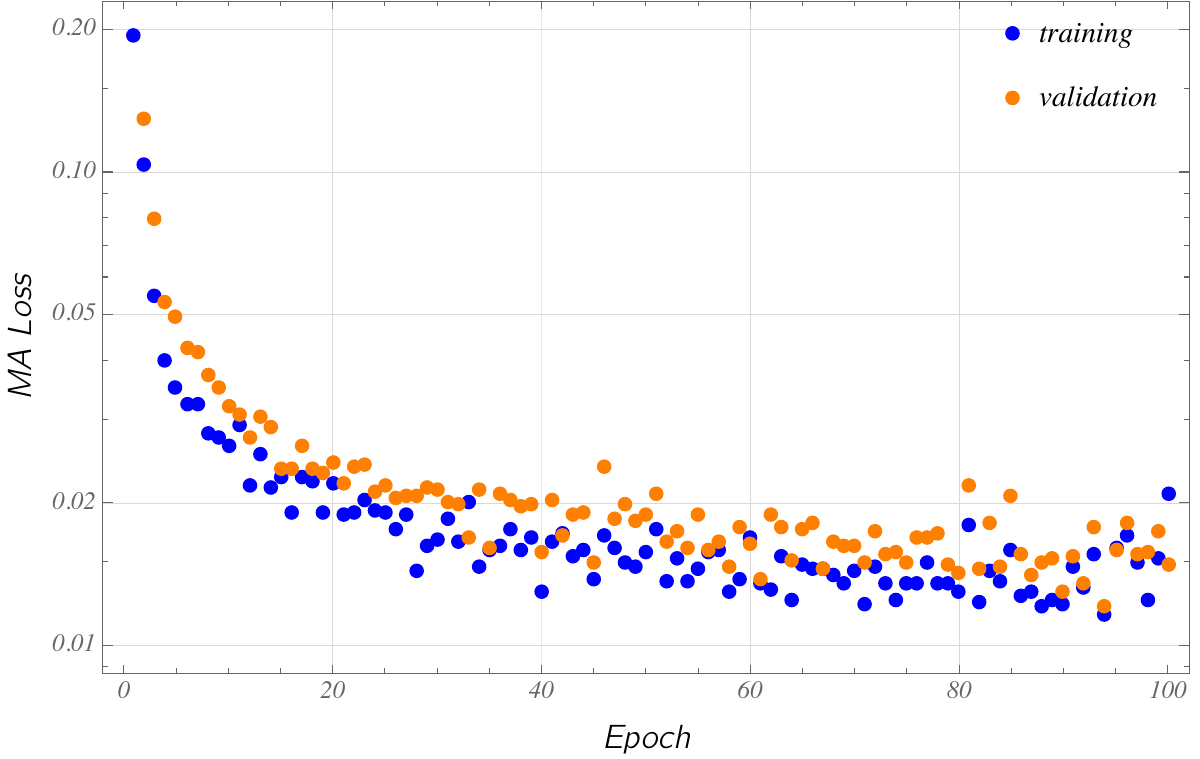}
    \caption{Monge-Amp\`ere loss for the neural network approximating the function $\phi$ defined in Eq.~\eqref{phidef}. For this example, we have chosen $\psi_0=2$ and $\psi=1$.}\label{fig:PhiLapLoss}
\end{figure}

\begin{figure}
    \centering
    \includegraphics[width=0.95\linewidth]{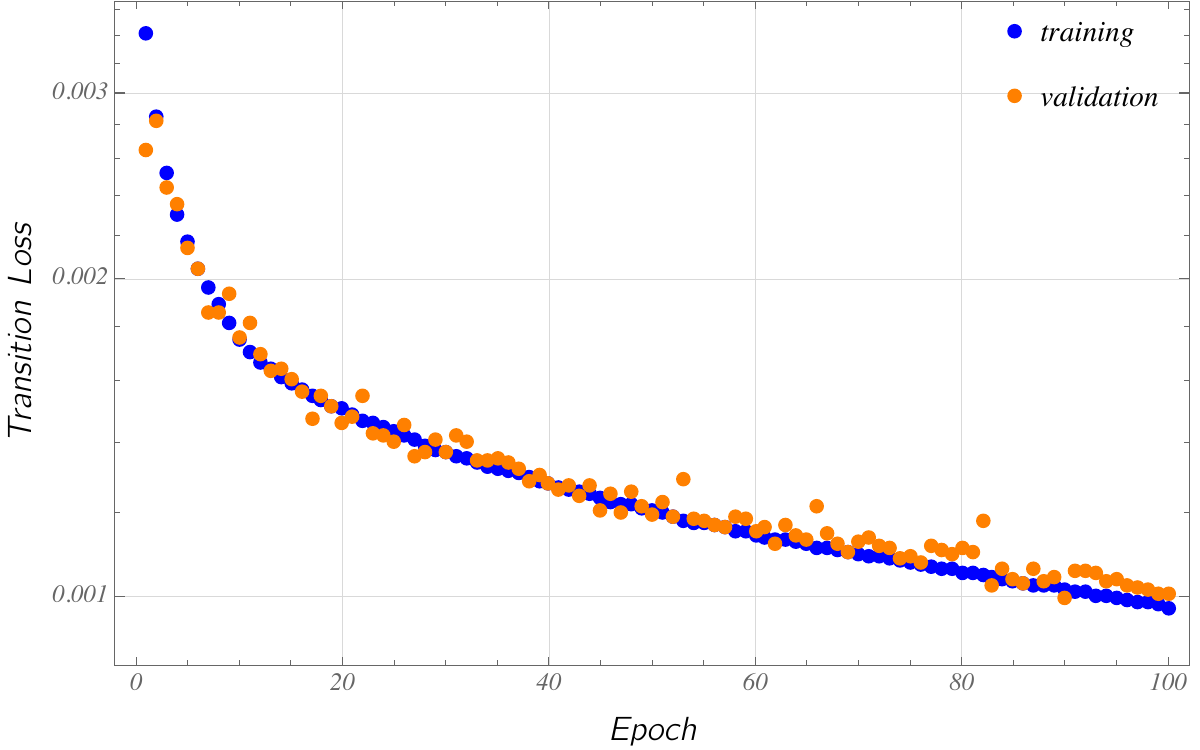}
    \caption{Transition loss for the neural network approximating the function $\phi$ defined in Eq.~\eqref{phidef}. For this example, we have chosen $\psi_0=2$ and $\psi=1$.}\label{fig:PhiTransLoss}
\end{figure}

The neural network is fully connected with GeLU activation \cite{hendrycks2023gaussian}, four layers and a width of 128. Training is carried out for 100 epochs, with batch size 64 and learning rate $0.001$. The change in loss over the course of a typical training round with $\psi_0=2$ and $\psi=1$ is illustrated in \fref{fig:PhiLapLoss} and \fref{fig:PhiTransLoss}. We define the following measures for the training performance,
\begin{align}
    M_{\text {MA}}[\phi] &= \frac{ \mathscr{L}_{\rm MA}[\phi]}{\mathscr{L}_{\rm MA}[\phi=0]},\\  \quad M_{\text {tr.}}[\phi] &= \frac{1}{\mathcal{V}}~\frac{\mathscr{L}_{\rm tr}[\phi]}{{\textrm{std.~dev.~of }} \phi}~,
\end{align}
where the integration is performed over the validation set. For a typical run with $\psi_0=2$ and $\psi=1$, we find $M_{\text {MA}}(\phi) = 0.04$ and $M_{\text {tr.}}(\phi) =0.05$.
These values indicate that the neural network $\phi$ performs well and leads to a good approximation of the Ricci-flat CY metric.

\subsection{The HYM bundle metric}\label{sec:beta}
\subsubsection{Mathematical background}
Let $\cL={\cal O}_X({\bf k})$ be a line bundle on $X$ associated with one of the matter fields in \eqref{spectrum}. To determine the correct field normalisations, knowledge of the HYM metric on ${\cal O}_X({\bf k})$ is required. As reference metric $H^{\FS}$, we choose the standard bundle metric associated with the Chern connection on ${\cal O}_\cA({\bf k})$ restricted to $X$. This can be written down explicitly as
\begin{equation}
 H^{\FS}=\left.\prod_{i=1}^4\kappa_i^{-k^i}\right|_X~.
\end{equation}
The HYM bundle metric $H$ on ${\cal O}_X({\bf k})$ is related to $H^{\FS}$ by
\begin{equation}\label{Hbeta}
 H=e^\beta H^{\FS}\; ,
\end{equation}
where $\beta$ is a real function on $X$. The HYM equation implies that $\beta$ must satisfy the following Poisson equation
%
\begin{equation}\label{betaeq}
    \Delta \beta = \rho_\beta = -g^{a\bar b} \partial_a\bar{\partial}_{\bar b}\ln\left(\bar H^{\FS}\right).
\end{equation}
The integrability condition for this equation amounts to the requirement that the slope of the line bundle vanishes, which is guaranteed by working at the supersymmetric K\"ahler locus in Eq.~\eqref{t_locus}. From the spectrum~\eqref{spectrum}, in order to determine the up-quark Yukawa couplings, we need to compute the three Hermitian bundle metrics on $\cL_2$, $\cL_5$ and $\cL_2^*\otimes\cL_5^*$.

\subsubsection{Computational realisation}
A possible method to solve Eq.~\eqref{betaeq}, which has, for instance, been used in Ref.~\cite{Larfors:2022nep}, is to construct a function basis, by starting with the sections of a given line bundle $\cL'\rightarrow X$. Then, Eq.~\eqref{betaeq} can be converted into a linear system by expanding $\beta$ in terms of this basis,
and by computing the matrix elements of the Laplacian as well as the components of the source $\rho_\beta$. In our case, the simplest line bundle for this purpose is $\cL'={\cal O}_X(1,1,1,1)$ with $16$ sections, leading to $16^2=256$ functions. Working with this basis requires computing $256^2$ Laplacian matrix elements. On the other hand, this is equivalent to working with the $l=0,1$ spherical harmonics in each of the $\mathbb{P}^1\cong S^2$ directions only, which indicates a rather poor approximation. 
Trying to improve this by starting with $\cL'={\cal O}_X(2,2,2,2)$ instead requires the calculation of an unfeasible number of matrix elements. In conclusion, it appears difficult to achieve sufficient accuracy with this method.

For this reason, we have opted to determine the HYM bundle metric by training a neural network, in analogy with what has been done for the CY metric. In practice, the neural network approximating the function $\beta$ in Eq.~\eqref{Hbeta} has two contributions for the loss function:  the HYM loss based on the failure to satisfy Eq.~\eqref{betaeq}, and the transition loss which ensures that $\beta$ transforms as a function. Explicitly, the loss function is given by
\begin{equation}\label{eqBetaLoss}
\begin{aligned}
\mathscr{L}  &= \alpha_1 \mathscr{L}_\text{HYM}+\alpha_2\mathscr{L}_\text{tr.}\\
\mathscr{L}_\text{HYM}[\beta]&=\ \big|\big|\Delta \beta - \rho_\beta \big|\big|_1\\
\mathscr{L}_\text{tr.}[\beta]&=\sum_{s\neq t} \big|\big|\beta_s-\beta_t\big|\big|_1\; ,
\end{aligned}
\end{equation}
where $\alpha_1$, $\alpha_2$ are weights for the two contributions, and
$\beta_s$, $\beta_t$ are the versions of $\beta$ computed on the patches $U_s$, $U_t$, respectively. 

\begin{figure}
    \centering
    \includegraphics[width=0.95\linewidth]{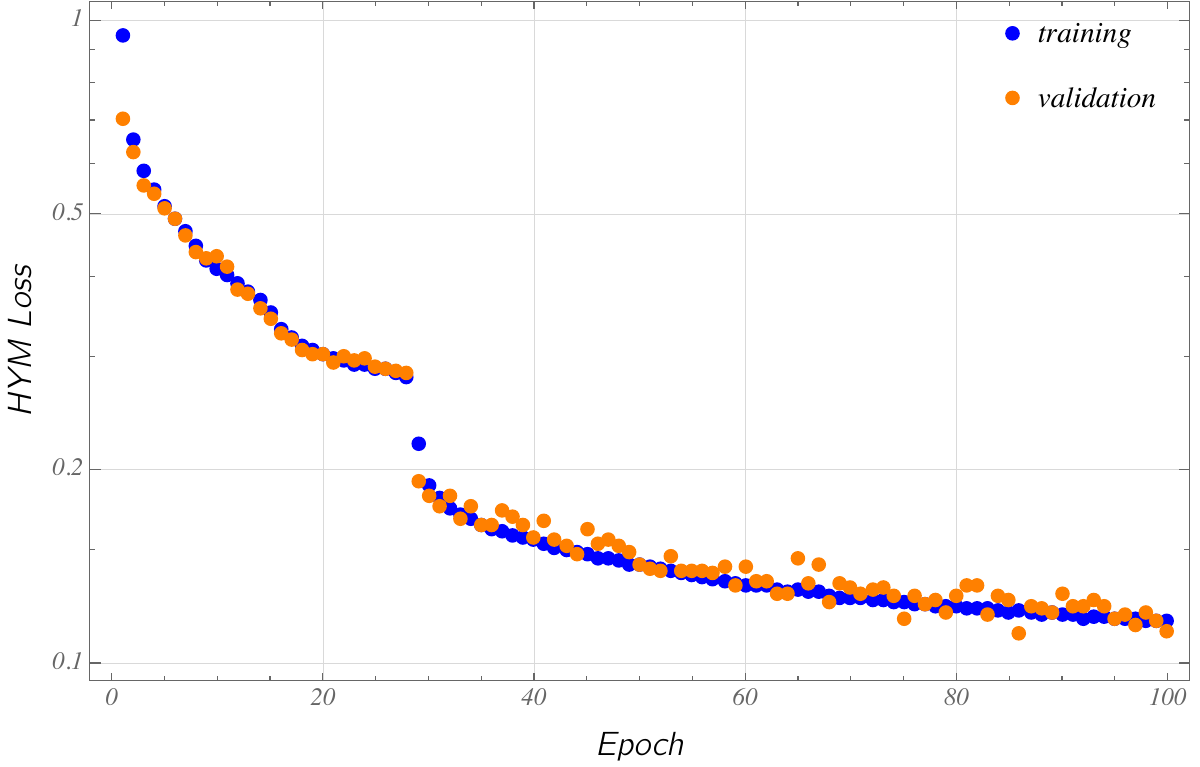}
    \caption{A typical HYM loss, $\mathscr{L}_{\rm{HYM}}$, as given in Eq.~\eqref{eqBetaLoss}, for the neural network approximating the function $\beta$ defined in Eq.~\eqref{betaeq}, and the line bundle $\mathcal{L}_2$. For this example, we have chosen $\psi_0=2$ and $\psi=1$.}
    \label{fig:LaplacianLossForBeta}
\end{figure}

\begin{figure}
    \centering
    \includegraphics[width=0.95\linewidth]{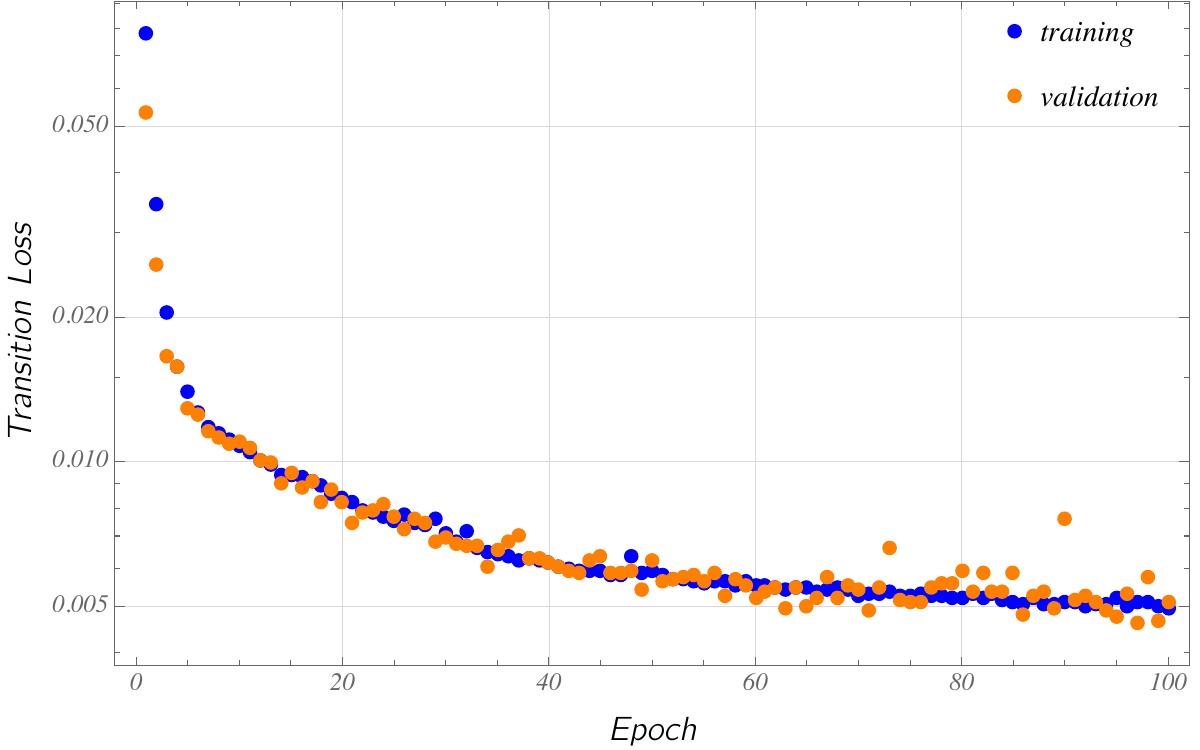}
    \caption{A typical transition loss, $\mathscr{L}_{\rm{transition}}$, as given in Eq.~\eqref{eqBetaLoss}, for the neural network approximating the function $\beta$ defined in Eq.~\eqref{betaeq}, for the line bundle $\mathcal{L}_2$. For this example, we have chosen $\psi_0=2$ and $\psi=1$.}
    \label{fig:TransitionLossForBeta}
\end{figure}

The neural network is fully connected with GeLU activation, three layers and a width of 128. Training is carried out for 100 epochs, with batch size 64 and learning rate $0.001$. Each of the three required bundle metrics is computed for $\psi_0=2$ and $\psi\in\{0,0.5,1,2,4,6\}$. To~discuss training and accuracy, we focus on the case $\psi=1$ which turns out to be typical. A typical change in loss over the course of training is shown in \fref{fig:LaplacianLossForBeta} and \fref{fig:TransitionLossForBeta}. Appropriate measures for the success or failure of training the neural networks can be defined by  
\begin{align}
    M_{\text {HYM}}[\beta] &= \frac{\mathscr{L}_{\rm HYM}[\beta]}{\int_X |\rho|},\\  \quad M_{\text {tr.}}[\beta] &= \frac{1}{\mathcal{V}} \frac{\mathscr{L}_{\rm tr.}[\beta]}{{\textrm{std. dev. of }} \beta}\; ,
\end{align}
where the integration goes over the validation set.

Typical numerical values for these measures, computed for the three line bundles, range between $0.05-0.06$ for $M_{\rm HYM}$ and between $0.02-0.04$ for~$M_{\rm tr.}$.  
We expect the product of the three bundle metrics to give the trivial bundle metric. It follows that this product must be constant over the CY manifold, and this is indeed true within an error of about 4\%.

\subsection{The harmonic forms}\label{sec:sigma}
\subsubsection{Mathematical background}
The matter fields in the spectrum~\eqref{spectrum} are associated with harmonic bundle-valued $(0,1)$-forms. Let $\cL$ be one of the relevant line bundles on $X$ with HYM bundle metric $H=e^\beta H^{\FS}$. Each matter field is represented by a specific cohomology class in $H^1(X,\cL)$ and we choose a reference form $\nu^{\FS}$ to represent this class. 
The unique harmonic form $\nu$ in the same cohomology class is related to $\nu^{\FS}$ by
\begin{equation}\label{nudef}
    \nu = \nu^{({\rm ref})} + \bar{\partial}_\cL \sigma\; .
\end{equation}
Here, $\sigma$ is a global section of $\cL$ determined by the Poisson equation
\begin{equation}\label{sigmaeq}
    \Delta_\cL \sigma = \rho_\sigma = - g^{a\bar b}\partial_a \left(H \nu_{\bar b}^{({\rm ref})}\right),
\end{equation}
where $\Delta_\cL$ is the Laplacian on $\cL$ relative to the Ricci-flat CY metric $g$ and the HYM bundle metric $H$.

For our model, we need to calculate seven harmonic forms, one for each of the matter fields in Table~\ref{tab:tabforHFmeasures}. These are precisely the forms which survive the $\Gamma=\mathbb{Z}_2\times\mathbb{Z}_2$ projection and the inclusion of the Wilson line. Explicit expressions for these reference forms have been given in Refs.~\cite{Blesneag:2015pvz,Blesneag:2016yag,Blesneag:2021wdf}, which, being somewhat lengthy, will not be reproduced here. 

\subsubsection{Computational realisation}
Following the same principles as before, we find the harmonic bundle-valued form $\nu$ from Eq.~\eqref{nudef}, by designing a neural network which approximates $\sigma$, trained on a loss function
with two contributions: the Laplacian loss, which measures the failure to satisfy Eq.~\eqref{sigmaeq}, and the transition loss, which ensures that $\sigma$ transforms as a section of $\cL$. In analogy with Eq.~\eqref{eqBetaLoss}, this loss function has the form
\begin{equation}\label{eqSigmaLoss}
    \begin{aligned}
        \mathscr{L}&=\tilde{\alpha}_1\mathscr{L}_{\Delta}+\tilde{\alpha}_2\mathscr{L}_{\rm tr.}\\
\mathscr{L}_{\Delta}[\sigma]&=\big|\big|\Delta_\cL\sigma-\rho_\sigma\big|\big|_1\\
\mathscr{L}_{\rm tr.}[\sigma]&=\sum_{s\neq t}\big|\big|\sigma_s-T_{(s,t)}\sigma_t\big|\big|_1\; ,
    \end{aligned}
\end{equation}
where $\tilde{\alpha}_1$, $\tilde{\alpha}_2$ are weights, $\sigma_s$, $\sigma_t$ are the versions of $\sigma$ computed on the patches $U_s$, $U_t$, respectively and $T_{(s,t)}$ denotes the transition function between $U_t$ and $U_s$.

Enforcing the right transformation for $\sigma$ is more complicated than in the previous cases for $\phi$ and $\beta$ due to the presence of the transition functions. These functions can become large or small away from the natural overlap region between two patches. As a result, for any $i\in\{1,\ldots,4\}$ with a line bundle integer $k^i\neq 0$, the transition loss is evaluated at points on a `belt' region $\mathcal B$, defined by $1/2<|z_i|<2$, where $z_i$ is the affine coordinate on the $i^{\rm th}$ $\mathbb{P}^1$ space introduced previously.

\begin{figure}
    \centering
    \includegraphics[width=0.95\linewidth]{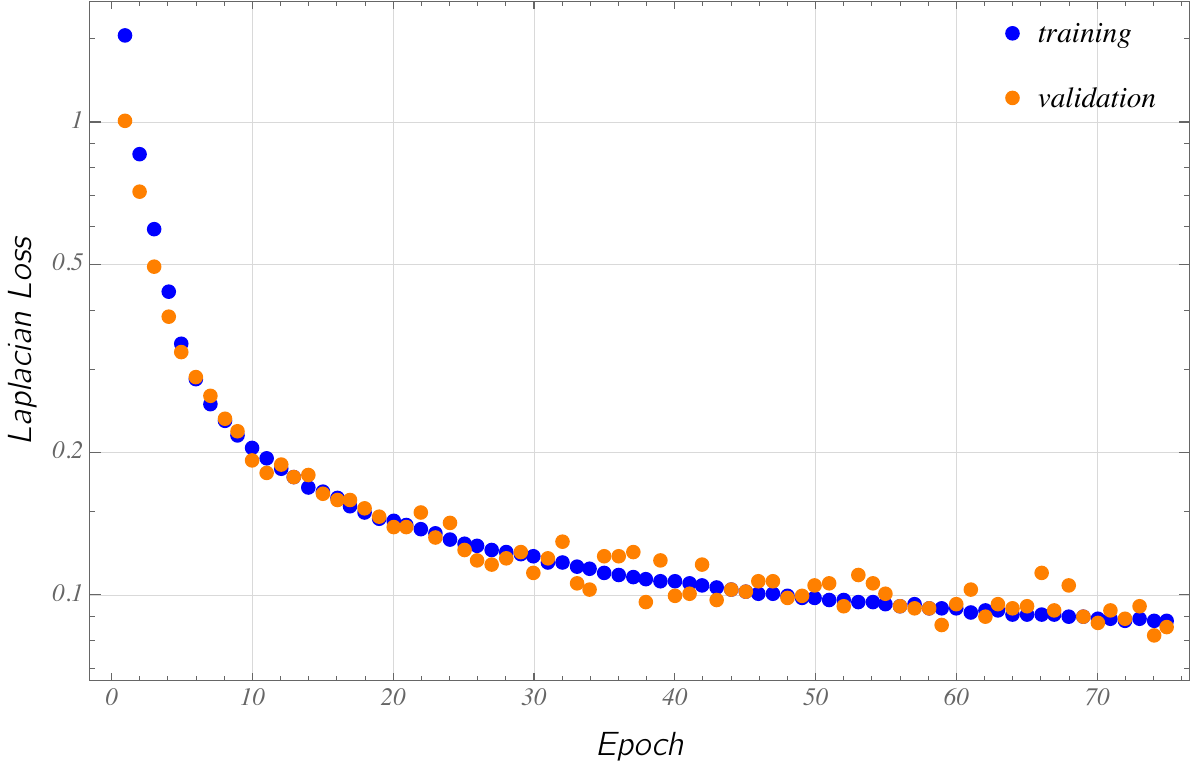}
    \caption{A typical Laplacian loss, $\mathscr{L}_{\Delta}$, as  in Eq.~\eqref{eqSigmaLoss}, for the neural network approximating the section $\sigma$ defined in Eq.~\eqref{sigmaeq}, for the case of the up-Higgs. For this example, we have chosen $\psi_0=2$ and $\psi=1$.}
    \label{fig:LaplacianLossForSigma}
\end{figure}

\begin{figure}
    \centering
    \includegraphics[width=0.95\linewidth]{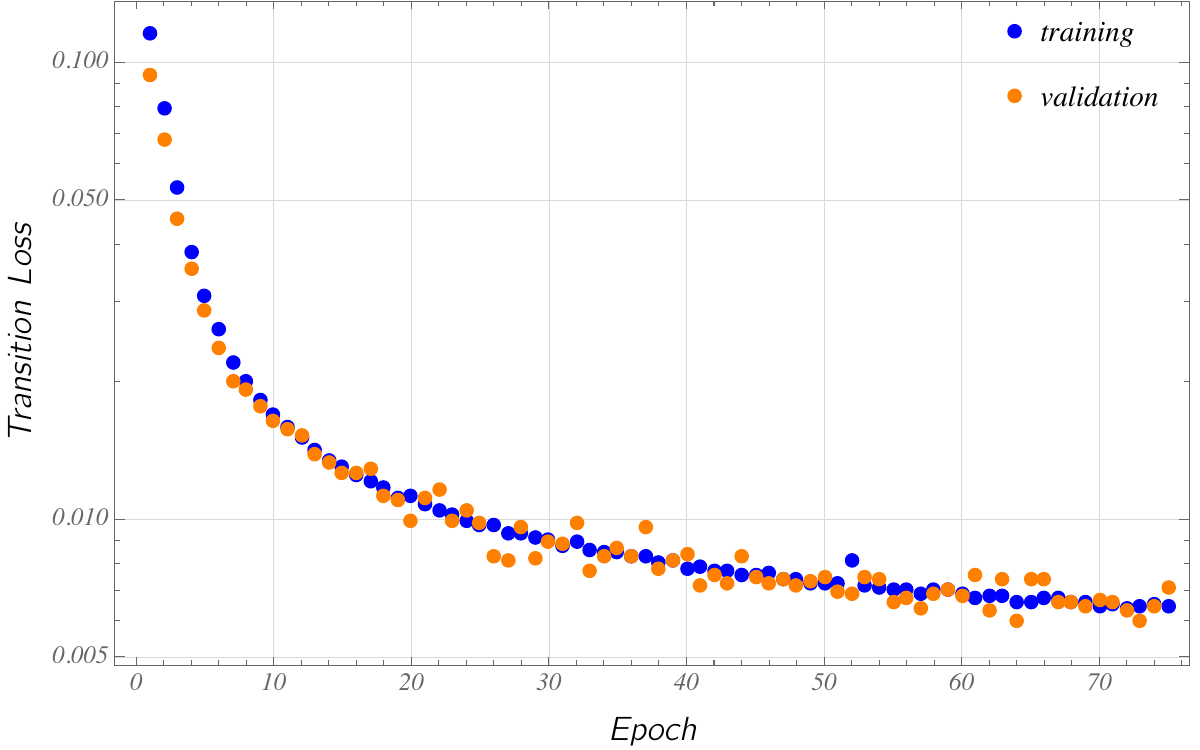}
    \caption{A typical transition loss, $\mathscr{L}_{\rm{tr.}}$, as given in Eq.~\eqref{eqSigmaLoss}, for the neural network approximating the section $\sigma$ defined in Eq.~\eqref{sigmaeq}, for the case of the up-Higgs. For this example, we have chosen $\psi_0=2$ and $\psi=1$.}
    \label{fig:TransitionLossForSigma}
\end{figure}

The neural network is fully connected with GeLU activation, three layers and a width of 128. Training is carried out for 75 epochs, with batch size 64 and learning rate $0.001$. Each of the required seven harmonic forms is computed for $\psi_0=2$ and $\psi\in\{0,0.5,1,2,4,6\}$. To~discuss training and accuracy, we focus on the case $\psi=1$ which turns out to be typical. The loss over the course of training for this case is shown in \fref{fig:LaplacianLossForSigma} and \fref{fig:TransitionLossForSigma}. Appropriate measures for the trained network's performance are defined as 
\begin{align}
    M_{\Delta}[\sigma] &= \frac{\mathscr{L}_{\Delta}[\sigma]}{\int_X |\rho|},\\  \quad M_{\text {tr.}}[\sigma] &= \frac{1}{\mathcal{V}_\mathcal B}~ \frac{\mathscr{L}_{\rm tr.}[\sigma]}{{\textrm{std. dev. of }} \sigma}\;,
\end{align}
where the integration goes over the validation set. In the evaluation of the transition measure for $\sigma$, we only consider the `belt' region with volume $\mathcal{V_B}$. Both measures are given in Table \ref{tab:tabforHFmeasures}, for each of the seven matter fields. 

\begin{table}[h]
    \centering
\begin{tabular}{|c|c|c|c|c|c|c|c|}
\hline 
\varstr{9pt}{4pt} &~$H^u_{2,5}$~& ~$Q_5$~  & ~$U_5$~  & ~$Q^1_2$~  & ~$Q^2_2$~  & ~$U^1_2$~  &~$U^2_2$~  \\\hline 
\varstr{9pt}{4pt} $~M_{\Delta}(\sigma)~$ & 0.06  & 0.08  & 0.12  & 0.15  & 0.20  & 0.14  & 0.12  \\\hline
\varstr{9pt}{4pt} $M_{\text {tr.}}(\sigma)$ & 0.04  & 0.09  & 0.13  & 0.15  & 0.17  & 0.21  & 0.14  \\
\hline
\end{tabular}
    \caption{Typical values of the errors, evaluated on the trained neural networks representing $\sigma$, for $\psi_0=2$ and $\psi=1$.}
    \label{tab:tabforHFmeasures}
\end{table}

We emphasise that, whilst this setup is very general, in the present case we are also able to construct neural networks which automatically obey the transition relations. This is done by choosing a set of (non-holomorphic) globally generating sections $\{\sigma_i,i=1,\cdots N\}$ of the bundle, that is, sections which do not simultaneously vanish anywhere on the CY manifold. A general section $\sigma$ is then written as 
\begin{equation}
    \sigma(x) = \sum_{i=1}^N f_i(x) \sigma_i,
\end{equation}
where $f_i$ are functions which are represented by the output of a neural network. Crucially these functions, along with the neural-networks used for $\phi$ and $\beta$, can be made invariant under projective transformations by making use of architectures analogous to those introduced in Ref.~\cite{Douglas:2020hpv}. As a result, $\sigma(x)$ automatically transforms as required between the different patches. Further details of these architectures will be explained in Ref.~\cite{futurePaper}.

\section{Yukawa couplings and masses}\label{sec:MassesMixings}
We now outline the main steps in the numerical calculation. For concreteness, we refer to the details of the model introduced in Section~\ref{sec:model}, but we note that the procedure is general for heterotic line bundle models.

\begin{enumerate}
\item[(1)] At the reference K\"ahler locus $(t_i)=(1,1,1,1)$, and for the defining polynomial~\eqref{pdef} 
with $\psi_0=2$ and for each $\psi\in\{0,0.5,1,2,4,6\}$, a sample of $300,000$ points on the TQ manifold is generated. Then, the Ricci-flat CY metric is computed by machine-learning the function $\phi$ in Eq.~\eqref{phidef}, with this point sample as a training/validation set.
\item[(2)] For each relevant line bundle $\cL$, the corresponding HYM bundle metric $H$ is computed by machine-learning the function $\beta$ in Eq.~\eqref{Hbeta}, with a newly generated training/validation point set and the loss function~\eqref{eqBetaLoss}. This needs to be done for three line bundles, namely, for $\cL_2$ associated to $Q_2^1,Q_2^2,U_2^1,U_2^2$, for $\cL_5$, associated to $Q_5,U_5$ and for $\cL_2^*\otimes \cL_5^*$, associated to $H^u_{2,5}$.
\item[(3)] For each matter field involved, the corresponding bundle-valued harmonic form $\nu$ is computed by starting with the corresponding reference bundle-valued form taken from Refs.~\cite{Blesneag:2015pvz,Blesneag:2016yag,Blesneag:2021wdf}, machine-learning $\sigma$ in Eq.~\eqref{nudef}, with an additional point sample as training/validation set and loss function~\eqref{eqSigmaLoss}. This needs to be done for seven cases: the three left-handed quarks $Q^i$, the three right-handed up-quarks $U^i$ and the up-Higgs $H^u$.
\item[(4)] Using the aforementioned quantities, Eqs.~\eqref{eqn:HolYuk} and \eqref{eqn:MFM} are used to compute the holomorphic Yukawa couplings and matter field metrics. This is done by Monte-Carlo integration using the given point sample. These results must be consistent with the structure of Yukawa couplings and matter field metrics as given in Eqs.~\eqref{Yhol} and \eqref{Kstruct_sec2}, which provides an important check of our calculation. Furthermore, this determines the entries $\lambda_i$ of the holomorphic Yukawa couplings in Eq.~\eqref{Yhol} and the quantities $\cK^Q,\cK^u,k^Q,k^u,k$ in Eq.~\eqref{Kstruct_sec2}.
\item[(5)] Finally, inserting these quantities into Eqs.~\eqref{Lphys}, \eqref{KP} and \eqref{masses}, we determine the physical up Yukawa couplings and up-quark masses.
\end{enumerate}
\begin{figure}
    \centering
    \includegraphics[width=0.95\linewidth]{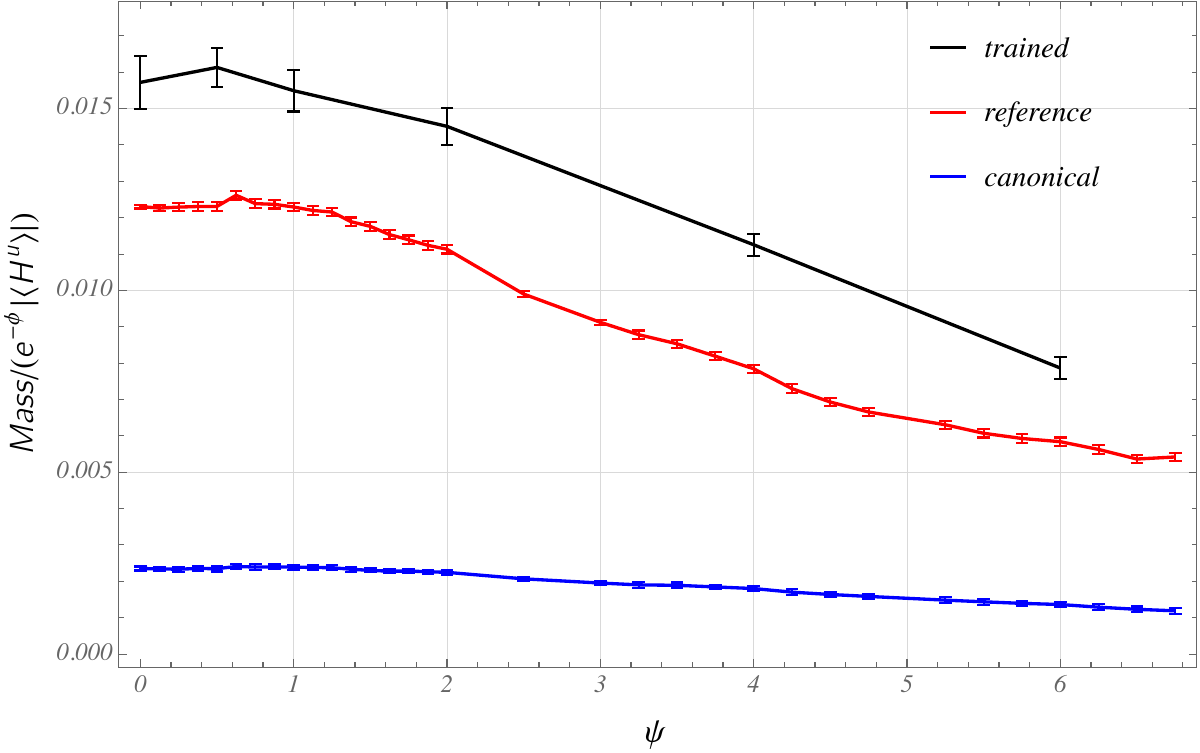}
    \caption{Plot of the (degenerate) top/charm mass in units of $e^{-\phi}|\langle H^u\rangle|$ for the defining polynomial~\eqref{pdef} with $\psi_0=2$, as a function of the complex structure modulus $\psi$. The black curve corresponds to the full neural network calculation, whilst the red curve gives the masses computed with the reference metrics and differential forms.  
    The blue curve shows the mass for canonical kinetic terms, obtained by setting $\mathcal{K}^Q = \mathcal{K}^u = \mathbb{I}_2$ and $k^u=k^Q=k=1$ in Eq.~\eqref{Kstruct_sec2}. Comparison of the blue and black curve demonstrates the importance of including the field normalisations. Error bars are statistical, and average five independent calculations.}
    \label{fig:mainresult}
\end{figure}
The above calculation is performed in three modes. Initially, we carry out a quick calculation with the analytic reference quantities, that is, we set  $\phi$ and all $\beta$ and $\sigma$ to zero. In this case, no neural networks need to be trained—the above first three steps are trivial—but the integrals still need to be carried out numerically. For comparison, we also calculate the masses which arise if one assumes canonical kinetic terms\footnote{Note that with this definition `canonical kinetic terms', the physical Yukawa couplings remain independent of the overall K\"ahler modulus.}, that is, by setting $\mathcal{K}^Q = \mathcal{K}^u = \mathbb{I}_2$ and $k^u=k^Q=k=1$ in Eq.~\eqref{Kstruct_sec2}. Finally, the results are then compared with the full calculation carried out by following all of the above five steps. In this case, a total of $11$ neural networks are trained to obtain the correct quantities $\phi$, $\beta$ and $\sigma$.

The final result for the up-quark mass is shown in \fref{fig:mainresult}, with the black curve showing the full calculation, the red curve the calculation based on the reference metrics and forms, and the blue curve the calculation with canonical kinetic terms. Due to the enhanced symmetry of the one-parameter family~\eqref{pdef}, the two non-zero masses are forced to be identical and, within numerical errors, this is confirmed by our results. This is the reason for why \fref{fig:mainresult} shows results for only a single mass. Some further remarks are in order. First, the physical mass has a significant dependence on complex structure (black curve). Secondly, comparison of the blue and black curves shows that taking into account the field normalisation has a significant effect. Finally, the calculation using the reference quantities (red curve) leads to a reasonable approximation, significantly better than the one based on assuming a canonical K\"ahler metric (blue curve). 

Can the degeneracy between the two non-zero masses be lifted if we move away from the one-parameter family~\eqref{pdef} of complex structures? To show that this is indeed possible, we consider the new defining polynomial in Eq.~\eqref{eqn:newPoly}. 
This point in moduli space was selected by randomly generating $5$ polynomials with the $\mathbb Z_2 \times \mathbb Z_2$ symmetry and integer coefficients between 0 and 100 and then taking the example with the largest mass ratio. As before, we carry out the calculation in three modes, that is, we perform the full calculation leading to up-quark masses $(m_i)$, the calculation with reference quantities leading to masses $(m^{\rm(ref)}_i)$ and the calculation with canonical kinetic terms with masses $(m^{\rm(can)}_i)$. The results are
\begin{equation}\begin{aligned}\label{eqn:liftDegMass}
    (m_1,m_2,m_3) &\approx
    e^{-\phi}|\langle H^u\rangle| \left(0, 0.009, 0.016\right)\;,\\
    (m^{\rm(ref)}_1,m^{\rm(ref)}_2, m^{\rm(ref)}_3)&\approx
    e^{-\phi}|\langle H^u\rangle| \left(0, 0.007, 0.013\right)\;,\\
    (m^{\rm(can)}_1,m^{\rm(can)}_2, m^{\rm(can)}_3)&\approx
    e^{-\phi}|\langle H^u\rangle| \left(0, 0.004, 0.008\right)\;.
\end{aligned}\end{equation}
This, as expected, lifts the degeneracy between the two non-zero masses.  The results lend further evidence to the claim that reference quantities serve as a better approximation than canonical kinetic terms. However, the larger of the two masses is still too small to account for the top mass and the split is not sufficient to explain the top to charm mass ratio. It is reasonable to expect that a more systematic exploration of the 20-parameter complex structure moduli space leads to masses which are more phenomenologically acceptable. This will be investigated in future work~\cite{futurePaper}.

\section{Conclusion}\label{sec:Conc}
In this paper, we have presented the first calculation of physical Yukawa couplings in a heterotic string model compactified on a CY manifold with non-standard embedding. In particular, we have calculated the physical up-quark Yukawa couplings and resulting masses. There are two main parts to our methodology. First, we have introduced analytic `reference' expressions for all relevant quantities which are contained in the right cohomology classes. Specifically, we have used the restricted ambient Fubini-Study metric as the reference CY metric, the restriction of the standard line bundle metrics on projective spaces as bundle reference metrics and certain restricted ambient bundle-valued forms as reference forms to represent the matters fields. In a second step, we have added exact terms to these reference quantities. For these we have determined numerical solutions using neural network techniques, in order to obtain the Ricci-flat CY metric, the Hermitian Yang-Mills bundle metrics and the harmonic bundle-valued forms. 

It turns out, the presence of additional $U(1)$ symmetries in our model leads to a rank two up-quark Yukawa matrix. The calculation has been carried out for the two-parameter family~\eqref{pdef} of tetra-quadric CY threefolds, whose additional symmetry forces the two non-zero masses to be equal. Our main result is shown in~\fref{fig:mainresult} which displays the value of the (degenerate) non-zero quark mass as a function of the complex structure parameter $\psi$ in Eq.~\eqref{pdef}. This is computed in three ways: (1)~the full numerical calculation (black curve), (2)~the semi-analytic calculation using the reference quantities (red curve), and (3)~a semi-analytic calculation where the kinetic terms have been set to canonical form `by hand' (blue curve). The full numerical calculation for each point in~ \fref{fig:mainresult} involves training 11 neural networks and takes about half a day on a twelve-core CPU. We estimate that the full numerical result is within 10\% of the actual values.

While the value of the mass and the degeneracy are clearly phenomenologically unacceptable, a number of interesting features can be observed in~\fref{fig:mainresult}. First, the mass has significant complex structure dependence, even for the limited two-parameter family under consideration (black curve). Secondly, the reference calculation (red curve) provides a reasonable approximation, within 25\% of the full numerical one (black curve), which is certainly better than the one based on canonical kinetic terms (blue curve). This observation might have important practical consequences for investigating the phenomenology of fermion masses from string theory. For a `first-pass' analysis, it may well be sufficient to calculate with the reference quantities. 

As emphasised earlier, the purpose of this paper is to provide proof of concept that a calculation of fermion masses can be carried out, by using machine learning techniques. Obtaining a realistic (up-quark) mass spectrum was not expected (and is indeed not achieved). Nevertheless, it is instructive to discuss the physical implications of our results for the model at hand. The fact that the mass matrix has, at the perturbative level considered here, one vanishing eigenvalue, is not necessarily a problem. Non-perturbative effects might well fill in some of the zero entries. Further problems are the degeneracy of the two non-zero masses and the fact, evident from~\fref{fig:mainresult}, that the value is too small to account for the top quark mass. 

To explore this further, we have moved away from the one-parameter family~\eqref{pdef}. Instead, we have carried out the calculation for a different point~\eqref{eqn:newPoly} in moduli space without any additional symmetry. The results in Eq.~\eqref{eqn:liftDegMass} show that the degeneracy between the two non-zero masses is indeed lifted, as expected. While this is encouraging, the actual values in Eq.~\eqref{eqn:liftDegMass} are still not phenomenologically viable.  Whether realistic top and charm quark masses can be obtained somewhere in the full $20$-dimensional moduli space remains to be investigated. These and other issues will be further studied in a forthcoming paper~\cite{futurePaper}, which will also present a systematic error analysis and more sophisticated network architectures which obviate the need for transition losses.

The present work opens up many directions for future investigation. The methods presented here apply to a wide range of CY manifolds, including complete intersections in products of projective spaces and hyper-surfaces in toric four-folds~\cite{He:2013ofa}. With suitable modifications, they should also be suitable for calculating masses in F-theory~\cite{Braun:2017feb}. In this paper, we have concentrated on vector bundles given as sum of line bundles. This leads to considerable technical simplifications, but also means that our techniques are directly applicable to the $\mathcal{O}(10^4)$ 
known line bundle models with the right particle content. Nevertheless, a generalisation to vector bundles with a non-Abelian structure group is feasible and desirable. Developing these techniques will open up the possibility for calculating fermion masses in large classes of string models and search for phenomenologically viable cases. Perhaps most crucially, this effort must ultimately be combined with moduli stabilisation~\cite{Anderson:2009nt, Anderson:2010mh, Anderson:2011cza, Anderson:2011ty,Deffayet:2023bpo}. 

\section*{Dedication}
This paper is dedicated to our colleague Graham G.~Ross who passed away in October 2021. Graham's many important contributions to high energy physics include his pioneering work on string phenomenology and fermion masses from string theory~\cite{greene:1986ar,greene:1986bm,greene:1986jb,greene:1987xh}. Some of the authors have greatly benefited from discussing these and related issues with Graham over the years, and these discussions have, in part, motivated and guided our work.\\[2mm] 

\section*{Acknowledgements}
AC was supported by a Stephen Hawking Fellowship, EPSRC grant EP/T016280/1, and by a Royal Society Dorothy Hodgkin Fellowship. CSFT is supported by the Gould-Watson Scholarship. TRH is supported by an STFC studentship. AL acknowledges support by the STFC consolidated grant ST/X000761/1. BO is supported in part by both the research grant DOE No.~DESC0007901 and SAS Account 020-0188-2-010202-6603-0338. 

The authors would like to thank Fabian Ruehle for support with {\sf cymetric}, Jonathan Patterson for assistance with the Oxford Theoretical Physics computing cluster Hydra, and Russell Jones for general technical support. BO would like to acknowledge the hospitality of the CCPP at New York University, where his contribution to this work was carried out.

\bibliography{bibliography}
\bibliographystyle{inspire}

\end{document}